\documentclass[doublecol]{epl2} 
\usepackage{graphicx} 
\usepackage{amsmath,amssymb,latexsym}
\usepackage{color}
\usepackage{footnote}
\usepackage{soul}
\usepackage{subcaption}

\title{On the impact of controlled wall Roughness Shape on the flow of a soft-material}
\shorttitle{Flowing of soft-materials in rough microchannels} 

\author{F. Pelusi\inst{1} \and M. Sbragaglia\inst{1} \and A. Scagliarini\inst{2} \and M. Lulli\inst{3} \and M. Bernaschi\inst{2} \and
  S. Succi\inst{4,2}}
\shortauthor{F. Pelusi \etal}

\institute{                    
\inst{1} Dipartimento di Fisica, Universit\`a di Roma  ``Tor Vergata'' and INFN - Via della Ricerca Scientifica 1, 00133  Rome, Italy
  \inst{2} Istituto per le Applicazioni del Calcolo, CNR - Via dei Taurini 19, 00185 Rome, Italy\\
  \inst{3} Department of Applied Physics, The Hong Kong Polytechnic University - Hung Hom, Kowloon, Hong Kong\\
  \inst{4} Center for Life Nano Science Sapienza, Istituto Italiano di Tecnologia - Viale Regina Elena 295, I-00161 Rome, Italy
}

\abstract{We explore the impact of geometrical corrugations on the near-wall flow properties of a soft-material driven in a confined rough microchannel. By means of numerical simulations, we perform a quantitative analysis of the relation between the flow rate $\Phi$ and the wall stress $\sigma_{\mbox{\tiny w}}$ for a number of setups, by changing both the roughness values as well as the roughness shape. Roughness suppresses the flow, with the existence of a characteristic value of $\sigma_{\mbox{\tiny w}}$ at which flow sets in. Just above the onset of flow, we quantitatively analyze the relation between $\Phi$ and $\sigma_{\mbox{\tiny w}}$. While for smooth walls a linear dependency is observed, steeper behaviours are found to set in by increasing wall roughness. The variation of the steepness, in turn, depends on the shape of the wall roughness, wherein gentle steepness changes are promoted by a variable space localization of the roughness.} 

\begin{document}

\maketitle

\section{Introduction}

\begin{figure*}
\begin{minipage}{1.0\textwidth}
\begin{center}
\onefigure[scale=0.68]{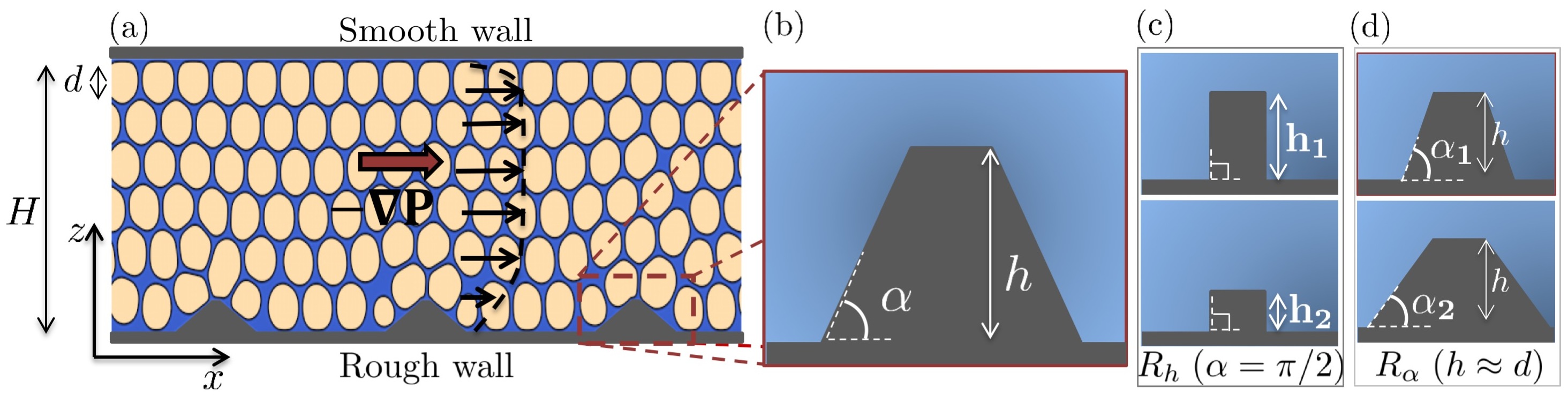}
\end{center}
\caption{Set-up for the numerical simulations of soft-material through rough microchannels. Using lattice-Boltzmann simulations (see~\cite{Scagliarinietal16,Derzsietal17,Derzsietal18} and references therein) we prepare droplets of a dispersed phase (light yellow) packed together in a continuous phase (dark blue) (Panel (a)). Droplet size distribution is slightly polydisperse with an average diameter $d$. The soft-material is confined in a channel of width $H \approx 7 d$ and driven with a constant pressure gradient along the stream-flow direction $x$. One wall of the channel comprises a regular roughness of equally spaced posts with trapezoidal shape (Panel (b)) with the height $h$ and the angle $\alpha$ being the free parameters in our analysis; other roughness parameters, such as the distance between two successive posts, remain fixed. We analyze two roughness realizations: by varying $h$ while keeping $\alpha=\pi/2$ ($R_{\mbox{\tiny h}}$, Panel (c)) and by varying $\alpha$ while keeping $h\approx d$ fixed ($R_{\alpha}$, Panel (d)).\label{fig:setup}}
\end{minipage}
\end{figure*}
Closely packed dispersions of soft particles are commonly encountered in industrial and every-day-life applications. Examples include concentrated emulsions, colloidal pastes, foams and gels~\cite{Larson95,Coussot05}. These ``soft-materials'' can sustain elastic deformations as solids do, and they are able to flow like non-Newtonian fluids for sufficiently large deformations, typically when the imposed stress exceeds a threshold value denoted as the ``yield stress''~\cite{Balmforth14,Bonn17}. This rheological complexity lies at the root of the widespread use of such materials across many disparate fields, such as cosmetics~\cite{Neves09}, food processing~\cite{Tabilo05}, pharmaceutical products~\cite{Islam04}, oil recovery~\cite{Chang99} and coatings~\cite{Maillard16} just to cite a few examples. Beyond the fundamental interest of understanding the microscopic mechanisms responsible for this rheological complexity~\cite{Balmforth14,Bonn17}, another important issue arises when these materials flow close to a solid wall. In such conditions, the bulk rheological complexity is inevitably coupled to boundary effects~\cite{Barnes95,Cloitreetal17}, and this coupling is more and more important at increasing confinement~\cite{Goyonetal08,Goyonetal10,MansardColin12}.
Typically, effective boundary conditions are introduced as a macroscopic manifestation of the microscopic interactions 
among soft particles and the wall. One of the most popular boundary effects is slippage~\cite{Neto05}. 
In the last two decades, there has been a considerable effort to understand the physical mechanisms responsible for the slip of soft-materials with complex rheology as recently reviewed in~\cite{Cloitreetal17}. For smooth walls, one needs to distinguish two different regimes~\cite{Meekeretal04a,Meekeretal04b,Sethetal08,Sethetal12}, depending on whether the wall stress is below the yield stress (``elastic'' regime) or above (``fluidized'' regime). In the elastic regime, it has been shown that the slip depends on the particle-walls interactions~\cite{Meekeretal04a,Meekeretal04b,Sethetal08,Sethetal12}. In the fluidized regime, the situation is further complicated because of the coupling between the bulk flow and wall phenomena~\cite{Goyonetal08,Goyonetal10}. In this landscape, various studies report scaling laws of the slip velocity as a function of the wall stress~\cite{Salmonetal03,Meekeretal04a,Meekeretal04b,Vayssadeetal14,Divouxetal15,Geraudetal13,Goyonetal10,Cloitreetal17,Sethetal12,PerezGonzalez12,Mansardetal14,Aktasetal14,Poumaereetal14,Joforeetal15,Ahonguio,OrtegaAvilaetal16}. Moreover, for soft-materials as foams, further enrichments are brought by the chemical properties of the surfactants used to stabilize the dispersion~\cite{Denkov05,Denkov06,Marze08}.\\ 
Another property that decidedly influences the near-wall properties is wall {\it roughness}. It is commonly accepted that the roughness suppresses the wall slip (see~\cite{Cloitreetal17} and references therein). Some authors also reported that wall slippage scales differently with the wall stress when using a rough wall instead of a smooth wall~\cite{Goyonetal10,Geraudetal13}. Moreover, there is evidence from different studies in the literature that roughness changes the {\it wall fluidization} of the soft-material close to the walls by promoting the emergence of plastic rearrangements due to the bumping of the soft particles against the wall asperities~\cite{Goyonetal08,Goyonetal10,MansardColin12,Geraudetal13,Mansardetal14,Paredes15,Scagliarinietal16,Derzsietal17,Derzsietal18}.  \\
In the emerging scenario, the flow shows complex near-wall properties where slippage and fluidization conspire in producing effective boundary conditions with different macroscopic behaviours. Systematic studies detailing how these effects relate to wall stresses for different roughness geometries are rare in the literature~\cite{Mansardetal14,Derzsietal17,Derzsietal18}. In this paper, 
we aim at taking a step further in this direction by presenting a comprehensive study on the impact of geometrical corrugations, 
by specifically focusing on the role of the roughness shape. We will concentrate mainly on wall stress values below -- or of the order of -- the material yield stress, mainly because for smooth walls the near-wall flow properties in this regime are understood in terms of elasto-hydrodynamics and particle-wall interactions~\cite{Meekeretal04a,Meekeretal04b,Sethetal08,Sethetal12,Cloitreetal17}, hence we can draw a parallel with the present results.
\section{Numerical method and simulations setup}
We resort to the lattice Boltzmann methodology (see~\cite{Scagliarinietal16,Derzsietal17,Derzsietal18} and references therein) to simulate a collection of droplets (see Fig.~\ref{fig:setup}) packed together in a continuous phase and driven by a constant pressure gradient applied in the stream-flow ($x$) direction of a confined channel. The code is a variant of the original implementation described in~\cite{Bernaschi09} that takes full advantage of the huge computing power of modern Graphics Processing Units (GPU), making possible to run the very large number of simulations required by this kind of systematic study in a reasonable time on a small cluster equipped with multiple GPUs. The packing fraction of the droplets is large enough to produce non-vanishing yield stress. The droplet size distribution is slightly polydisperse with an average diameter $d$. Wetting properties are chosen in such a way that the droplets do not adhere to the walls, so as to fully appreciate the role of roughness~\cite{Scagliarinietal16,Derzsietal17,Derzsietal18}. Periodic boundary conditions are applied in the stream-flow direction, whereas two walls confine the system along the $z$ direction: one wall ($z=H$) of the channel is smooth, whereas the other ($z=0$) is patterned with a periodic roughness comprising equally spaced posts of trapezoidal shape with the height $h$ and the angle $\alpha$ being the free parameters in our study (see Fig.~\ref{fig:setup}). 
We also performed simulations with both smooth walls, to compare with the rough cases. The distance between the posts is kept 
fixed to $\lambda=8 d$. Previous studies investigated~\cite{Derzsietal17,Derzsietal18} the impact of a change in 
$\lambda$ at fixed post shape; we are interested, on the countrary, in assessing the impact of different shapes of the posts. 
To that purpose, we analyze two different realizations of the {\it roughness shape}: i) ``varying-h'' roughness with $\alpha$ fixed and equal to $\pi/2$ (roughness realization $R_{\mbox{\tiny h}}$); ii) ``varying-$\alpha$'' roughness keeping $h$ fixed to $h \approx d$ (roughness realization $R_{\alpha}$). As a matter of fact, the roughness realization $R_{\mbox{\tiny h}}$ is the most widely used and studied in the literature~\cite{Goyonetal08,Mansardetal14,Scagliarinietal16,Derzsietal17,Derzsietal18}. However, by thinking of a generic roughness profile $h(x)$, one could say that in the realization $R_{\mbox{\tiny h}}$ the roughness results from different height variations ``strongly'' localized in space; hence, a possible alternative situation is that of fixed height with ``variable localization'' in space, which motivates the choice of the realization $R_{\alpha}$. Moreover, the $R_{\alpha}$ realization enjoys the 
interesting feature that, for a particular angle, it can optimally ``accommodate'' the hexagonal arrangement of 
dense monodisperse emulsions in $2d$, therefore allowing for the assessment of possible effects on the rheology coming 
from such ``topological affinity''. Summarizing, the two roughness realizations are chosen in a complementary way so as to highlight both the importance of height variations as well as the roughness localization in space.  Notice that in both realizations the roughness value $R$ is obtained as the ratio between the real area of the rough wall and the projected one, so it is equal to unity for smooth walls. $R$ increases with the height $h$ in the realization $R_{\mbox{\tiny h}}$ and with $\alpha$ in the realization $R_{\alpha}$. From the methodological point of view, with respect to our previous investigations~\cite{Scagliarinietal16,Derzsietal17,Derzsietal18}, the change in roughness shape is the added value brought by the present study. Given the roughness realization and value, the control parameter for the simulations is the pressure gradient, which is tuned so as to produce a value of wall stress $\sigma_{\mbox{\tiny w}}$ {below - or of the order of - the material yield stress. The wall stress is an outcome of the simulations and can only be measured a posteriori~\cite{Belardinelli13,ourJFM15}, once the simulation has reached a steady state. We remark that by wall stress $\sigma_{\mbox{\tiny w}}$ we indicate the stress in contact with the rough wall \footnote{When we perform simulations where both walls are smooth, the stress profile is symmetric and we take $\sigma_{\mbox{\tiny w}}$ to be the stress in contact with one of the two walls.}: this is the physical observable required to obtain a fair assessment of the role that different roughness shapes play in the resulting near-wall flow properties. To the purpose of running many simulations for different roughness shapes/values and wall stress values, we kept the wall-to-wall resolution fixed to $H \approx 7 d$. All the results of the numerical simulations are reported in lattice Boltzmann units, hereafter lbu. \\
\begin{figure}[t!]
\onefigure[scale=0.15]{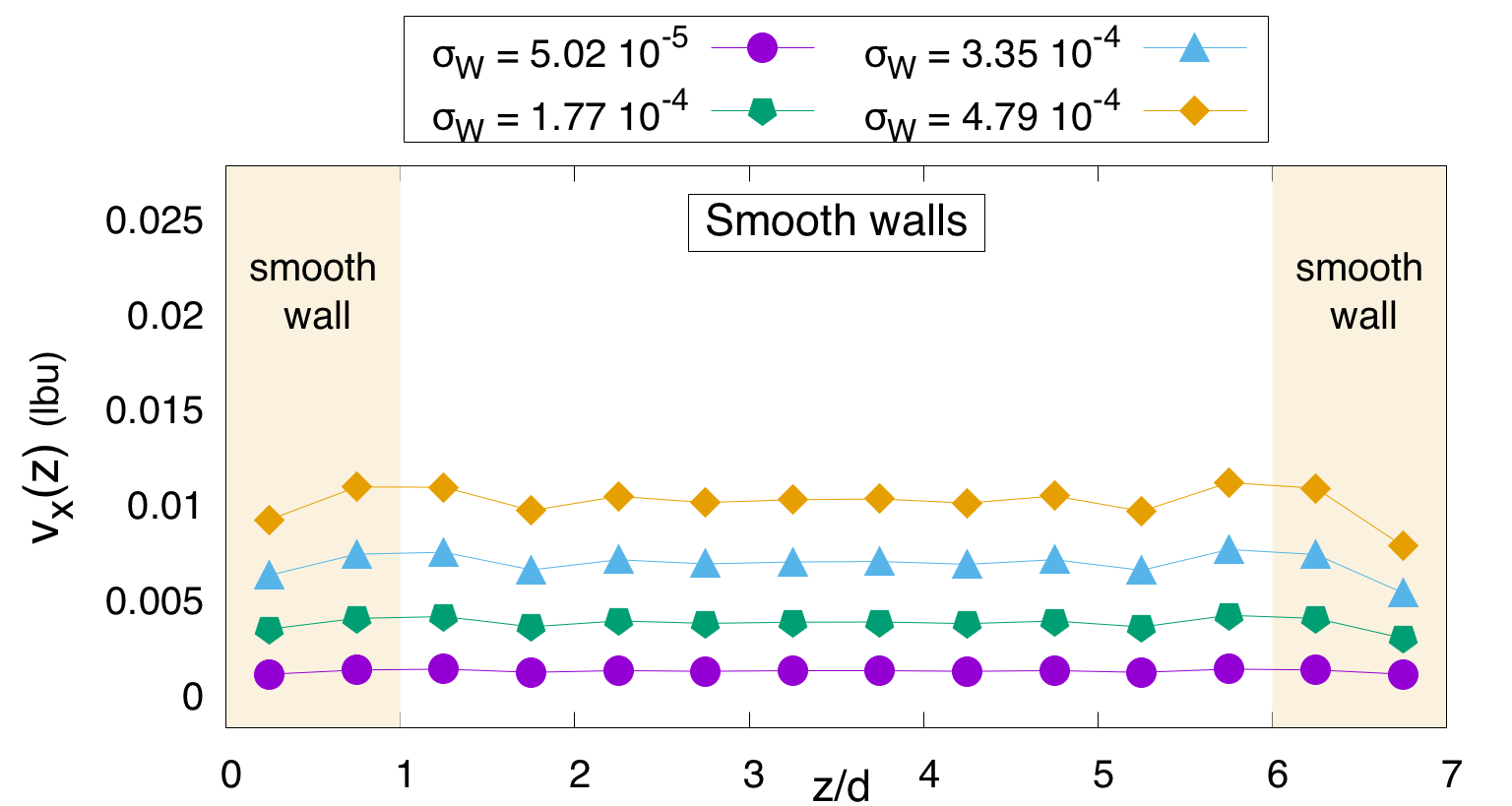}
\onefigure[scale=0.15]{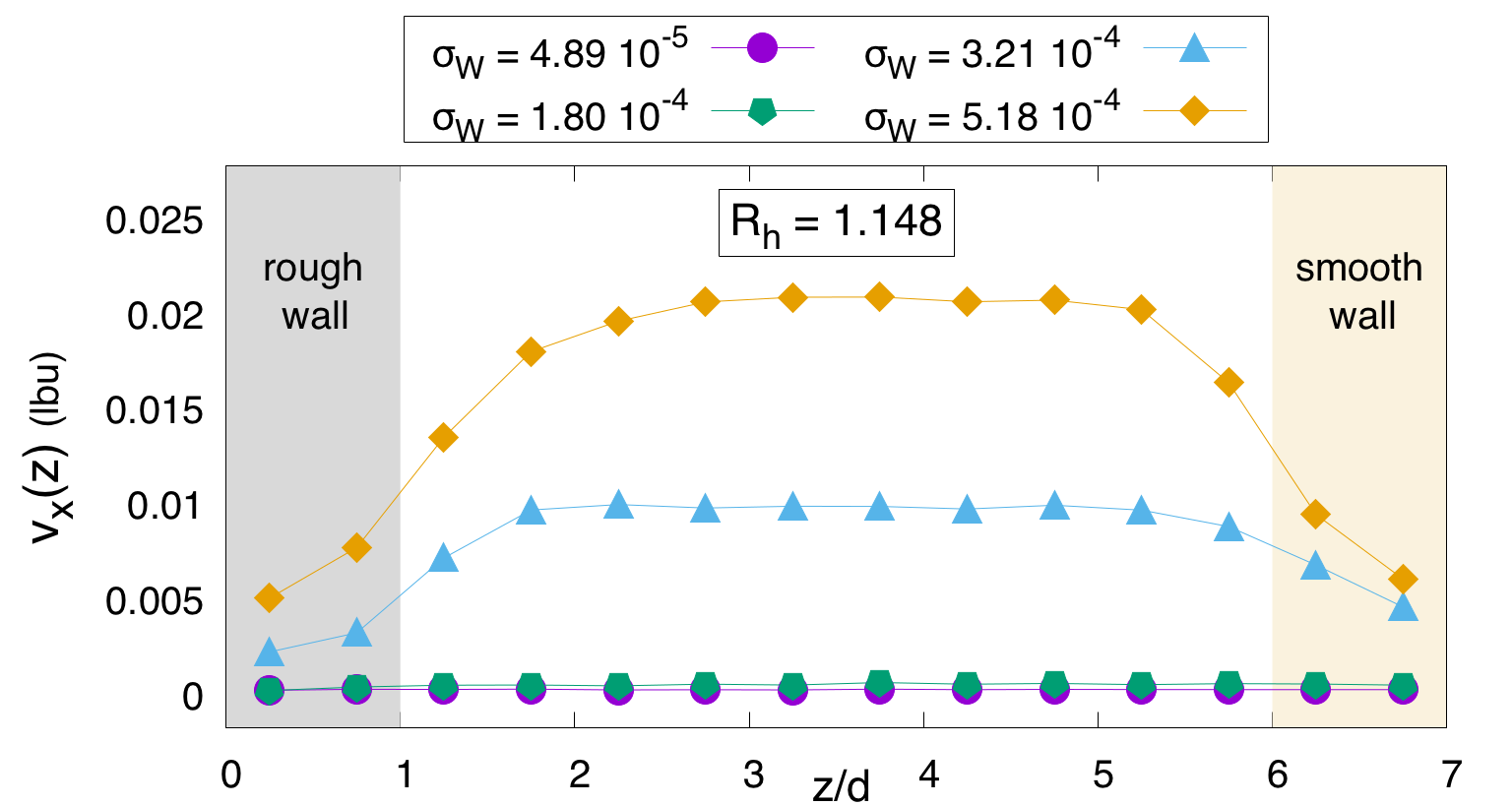}
\onefigure[scale=0.15]{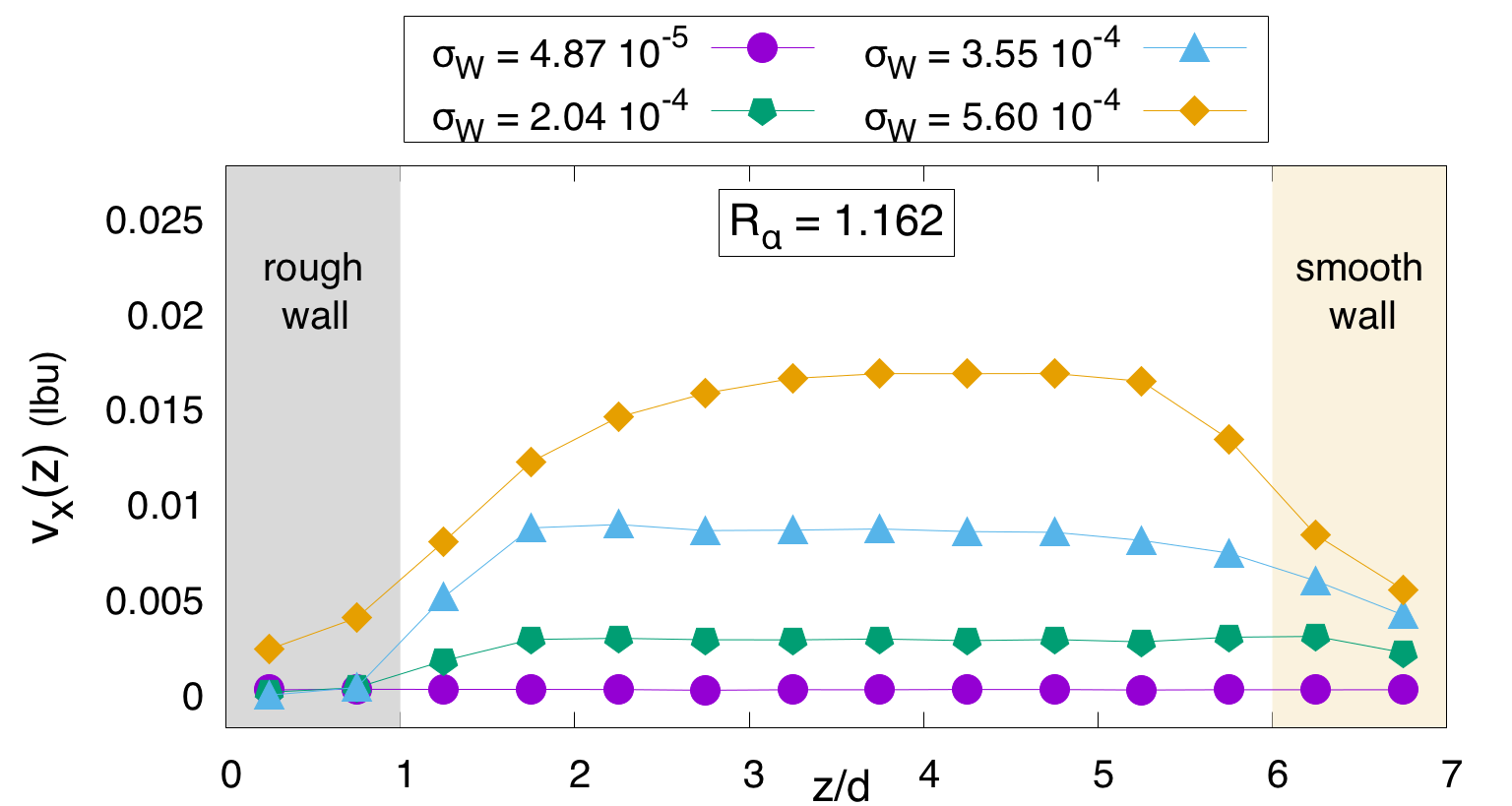}
\caption{Velocity profiles for different values of the wall stress $\sigma_{\mbox{\tiny w}}$. Top panel: channel with both smooth walls. Central panel: rough channel with realization $R_{\mbox{\tiny h}}$ ($R_{\mbox{\tiny h}}=1.148$). Bottom panel: rough channel with realization $R_{\alpha}$ ($R_{\alpha}=1.162$).}
\label{fig:veloprofiles}
\end{figure}
\begin{figure}[t!]
\onefigure[scale=0.15]{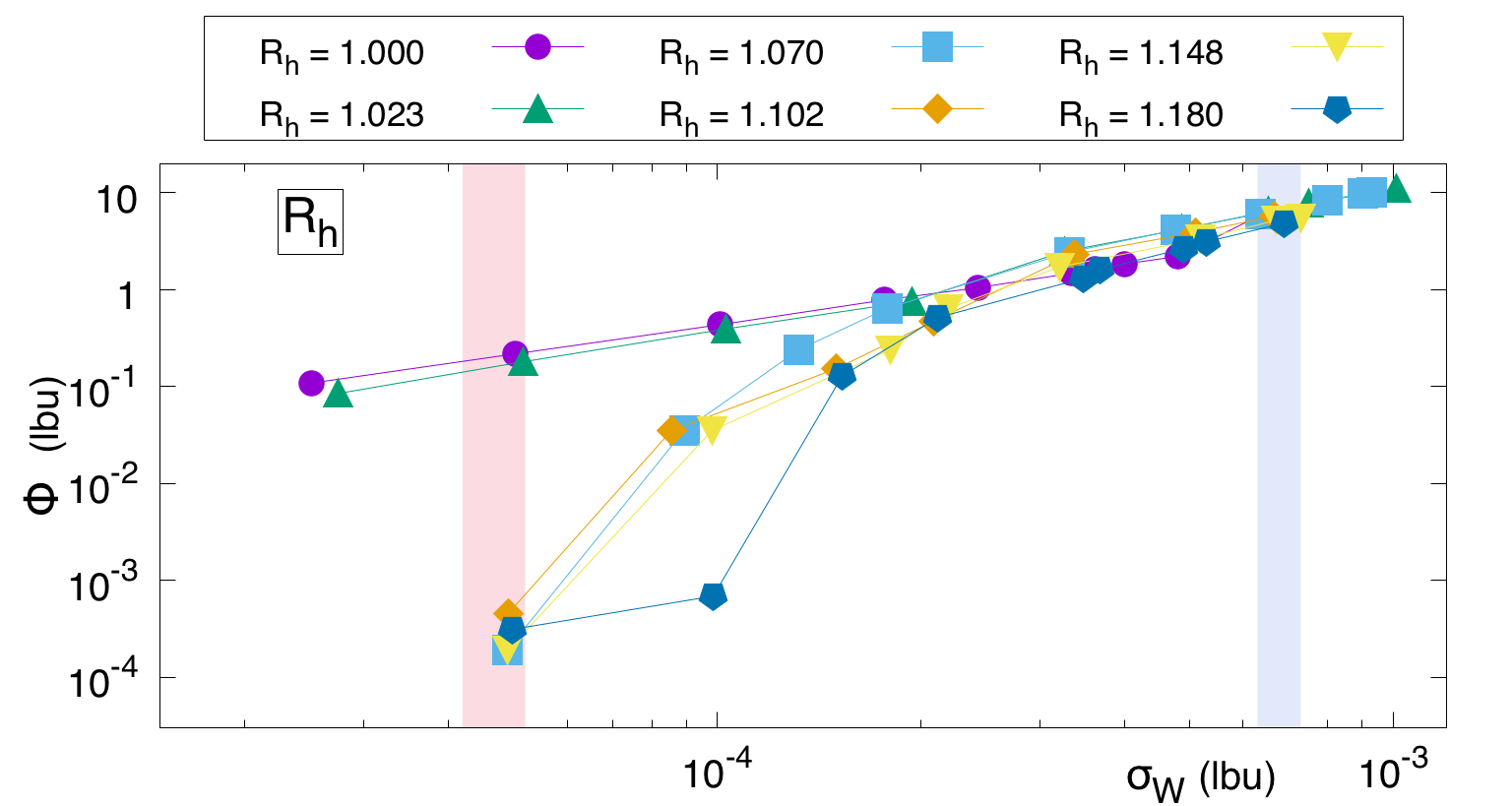}
\onefigure[scale=0.15]{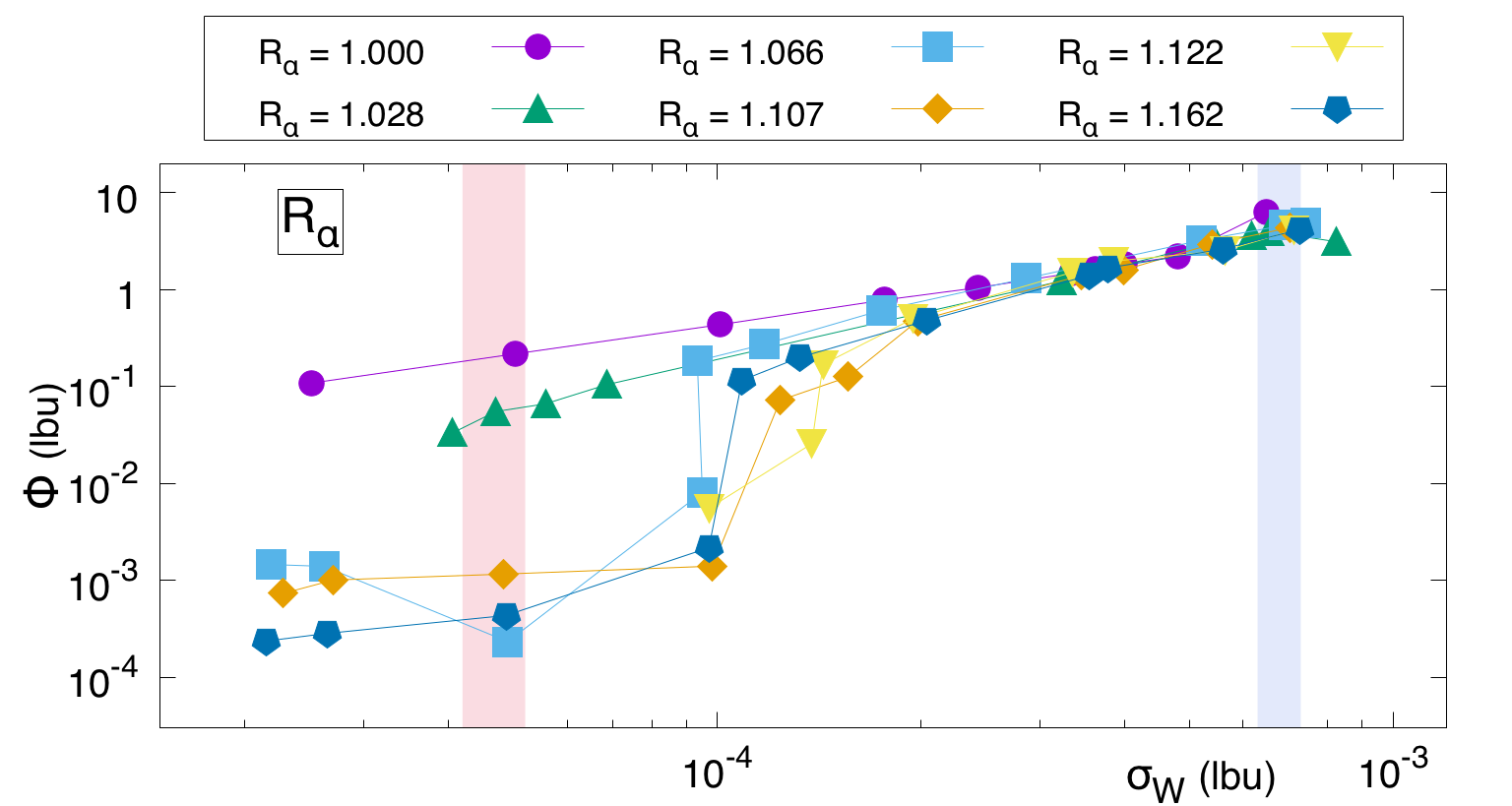}
\caption{Log-log plot of the flow rate $\Phi$ as a function of the wall stress $\sigma_{\mbox{\tiny w}}$ for two different realizations of the roughness shape: $R_{\mbox{\tiny h}}$ (top panel) and $R_{\alpha}$ (bottom panel). Shaded regions refer to two specific wall stress intervals later analyzed in Fig.~\ref{fig:flowRate-R}.}
\label{fig:scaling}
\end{figure}
%
\section{Results}
We average the local instantaneous velocity along the stream-flow coordinate, $x$. 
The resulting velocity profiles, $v_x(z,t)$, are further integrated in $z$ to obtain the instantaneous flow rate $\phi(t)$. 
Both $\phi(t)$ and $v_x(z,t)$ are, in general, fluctuating quantities; thus, we define the flow rate $\Phi$ and 
the velocity profile $v_x(z)$ as the time average over the (statistically) steady state of $\phi(t)$ and $v_x(z,t)$. 
Some velocity profiles $v_x(z)$ for the smooth channels ($R_{\mbox{\tiny h}}=R_{\alpha}=1.0$) and the rough channels ($R_{\mbox{\tiny h}}=1.148$ and $R_{\alpha}=1.162$) are displayed in Fig.~\ref{fig:veloprofiles} as a function of $z/d$. Since all the values of the 
wall stresses are below -- or of the order of -- the yield stress of the soft-material, the velocity profiles essentially display a ``plug dominance'' in which the wall slip (or effective slip) is the relevant contribution to the flow rate. When the values of 
$\sigma_{\mbox{\tiny w}}$ increase, the velocity profiles in the rough channels display a bending, strongly localized close to 
the walls, in a region of width comparable to the mean droplet diameter. 
Physically, these changes in the velocity profiles are generated by {\it boundary yielding events} of droplets of the bulk phase 
in the near-wall region. Overall, we observe that roughness (central and bottom panels) has the tendency to promote yielding 
events at wall stresses for which yielding events are not observed for the channel with both smooth walls (top panel). 
Additionally, such boundary yielding is asymmetric~\cite{Derzsietal17,Derzsietal18}, with the yielding close to the rough wall ($z=0$) being more ``vigorous'' with respect to the smooth wall case ($z/d=7$). 
Given this scenario, we characterized the dependence of the flow rate $\Phi$ on the wall stress $\sigma_{\mbox{\tiny w}}$. 
This is shown in Fig.~\ref{fig:scaling}. This figure essentially highlights that roughness generically suppresses the flow 
at small $\sigma_{\mbox{\tiny w}}$ (no-flowing regime) and also that there exists a particular value of $\sigma_{\mbox{\tiny w}}$ 
at which the system transits from a no-flowing regime to a flowing one. The crossover between the two regimes signals the 
presence of yielding events strongly localized close to the boundary, which are clearly visible in the velocity profiles 
(see Fig.~\ref{fig:veloprofiles}). At a closer look, we notice that for the smaller roughness values in the realization 
$R_{\alpha}$ the crossover is ``preceded'' with respect to the realization $R_{\mbox{\tiny h}}$, 
pointing to the fact that for fixed $\sigma_{\mbox{\tiny w}}$ the movement of the material in the near-wall region is facilitated 
by a variable localization in space. An important issue pertains to the quantitative relation between the flow rate $\Phi$ 
and the wall stress $\sigma_{\mbox{\tiny w}}$~\cite{Meekeretal04a,Meekeretal04b,Sethetal08,Sethetal12,Cloitreetal17}. 
To investigate this point and to provide quantitative details on the simulations with rough walls in comparison to those 
with smooth walls, we proceeded as follows. First of all, we focused {\it only} on the flowing regime, i.e. we excluded those $\sigma_{\mbox{\tiny w}}$ at which the flow is practically suppressed by the rough walls. To ease the comparison of the curves, we selected a reference wall stress $\sigma_{\mbox{\tiny w}}^{(0)}$ at the beginning of the flowing regime and we identified the corresponding reference flow rate $\Phi^{(0)}$ (the corresponding values have been summarized in Table~\ref{table}). We have then analyzed the relation between the normalized flow rate $\Phi/\Phi^{(0)}$ and the normalized wall stress $\sigma_{\mbox{\tiny w}}/\sigma_{\mbox{\tiny w}}^{(0)}$. Results are reported in Fig.~\ref{fig:scalingZoom}. For channels with smooth walls ($R_{\mbox{\tiny h}}=R_{\alpha}=1.0$), the relation is essentially linear. We also observe that steeper curves set in whenever we increase the value of wall roughness. Following the literature~\cite{Goyonetal10,Geraudetal13,Mansardetal14}, we quantified such steeper behaviours via the scaling relation $\Phi/\Phi^{(0)} = (\sigma_{\mbox{\tiny w}}/\sigma_{\mbox{\tiny w}}^{(0)})^p$ with $p \ge 1$, where the value of $p$ is given in the insets of Fig.~\ref{fig:scalingZoom}. In both roughness realizations, it is observed an increase of $p$ when changing from smooth channels to rough channels, with the value of $p$ increasing as roughness value increases. In the literature, the change of exponent from 1 to larger values has been well substantiated for droplets moving close to smooth walls. In particular, a linear scaling between slippage and wall stress is observed when droplets do not adhere to the walls; a quadratic scaling, instead, is observed when droplets adhere to the walls and can be understood as the result of a balancing between flow-induced deformation and elastic resistance (i.e. elasto-hydrodynamics)~\cite{Meekeretal04a,Meekeretal04b,Sethetal08,Sethetal12}. The linear scaling law observed in our simulations with smooth walls is therefore plausible, because we are dealing with droplets that do not adhere to the walls and we are essentially in a regime where the flow profile is a plug (i.e. dominated by wall slip, see top panel Fig.~\ref{fig:veloprofiles}). Regarding the increase of the exponent for the rough walls, one could naively think that the roughness induces deformations in the droplets and these deformations are opposed by the elasticity of the droplets themselves; in other words, one could point to the same ingredients that predict a quadratic scaling in elasto-hydrodynamic models~\cite{Meekeretal04a,Meekeretal04b,Sethetal08,Sethetal12}. Notice that previous experimental findings of Goyon {\it et al.}~\cite{Goyonetal10} and Geraud {\it et al.}~\cite{Geraudetal13} actually report quadratic scaling laws for emulsion/carbopol gels flowing close to rough walls, as opposed to linear scaling laws observed for smooth walls. The added value brought by the present numerical simulations is that we can quantitatively explore how such steeper curves set in, starting from the smooth walls and increasing the roughness values via two independent ``pathways'' (realizations $R_{\mbox{\tiny h}}$ and $R_{\alpha}$). Interestingly, we find that the roughness shape impacts the way the exponent $p$ changes with the roughness, with the pathway $R_{\alpha}$ allowing for more gentle variations, especially at small roughness values.}\\ 
While the $\Phi$ - $\sigma_{\mbox{\tiny w}}$ relation with rough walls is manifestly steeper than the one with the smooth walls, 
some comments on the scaling scenario advocated before are in order. 
Although the range of $\sigma_{\mbox{\tiny w}}$ over which the exponent $p$ can be estimated is relatively narrow 
(roughly two decades for smooth walls and less than one decade for rough walls (see Fig.~\ref{fig:scaling})), 
it should be remarked that this is in the order found in experiments as 
well~\cite{Goyonetal10,Sethetal12,Geraudetal13,Mansardetal14,Salmonetal03,PerezGonzalez12,Vayssadeetal14,Aktasetal14,Joforeetal15}. 
As a matter of fact, in line with the ``soft'' nature of the material,
    physical bounds to the achievable stresses emerge: on one extremum there is the yield stress (below which there is no flow) 
and on the other there is the film pressure between droplets, above which coalescence occurs. 
Both such stresses typically do not differ much from
    each other. Nevertheless, more refined interface models may improve the emulsion stability~\cite{Montessori19} and enlarge, then,
    the scaling range, this making a subject of decided interest for future studies.\\
Finally, in Fig.~\ref{fig:flowRate-R} we report the results for $\Phi$ as a function of the roughness value, for both roughness realizations and for the two intervals of $\sigma_{\mbox{\tiny w}}$ indicated in Fig.~\ref{fig:scaling}. Data for small $\sigma_{\mbox{\tiny w}}$ (top panel) reveal that the drop to zero of the flow rate is practically continuous; close to the critical roughness -- where $\Phi$ becomes essentially zero -- both realizations yield comparable values. Notice that the values of $\sigma_{\mbox{\tiny w}}$ considered in the top panel correspond to a situation where the material either advances as a plug or stops, i.e., no boundary yielding events are present. For large values of $\sigma_{\mbox{\tiny w}}$ (bottom panel), instead, a clear quantitative distinction between the two roughness realizations is observed: while $\Phi$ for $R_{\mbox{\tiny h}}$ is monotonously decreasing, $\Phi$ for $R_{\alpha}$ appears rather constant. This tells us that in presence of boundary yielding events, the blockage effect is mainly driven by the roughness height. 
Incidentally, we remark that Mansard {\it et al} \cite{Mansardetal14} found that the flow rate increased again with
the post height (increasing $R_{\mbox{\tiny h}}$ in our language) as the latter reached a value comparable with the average 
droplet size. They argued that this phenomenon could be due to the fact that roughness of height comparable 
with the droplet size contributes to restoring configurational order next to the boundary.
This might explain why we do not observe such non-monotonic behaviour: our emulsions, in fact, have a much lower
degree of polydispersity; therefore the crystalline order is already a built-in property
of the system, that the roughness tends to disrupt, rather than restore. As a matter of fact, the non-trivial interplay between
wall properties and polydispersity is far from being understood and provides a strong motivation 
for further studies in this direction.

\begin{table}[t!]
  \begin{minipage}[b]{1.\linewidth}\centering
    \begin{tabular}{|c|c|c|}
      \hline
      $R_{\mbox{\tiny h}}$ & $\Phi^{(0)}$ (lbu) & $\sigma_{\mbox{\tiny w}}^{(0)}$ (lbu) \\
      \hline
      $1.000$ & $0.434$  & $1.01\;10^{-4}$   \\
      $1.008$ & $2.592$  & $3.19\;10^{-4}$   \\
      $1.023$ & $2.494$ & $3.26\;10^{-4}$   \\
      $1.070$ & $2.442$ & $3.31\;10^{-4}$   \\
      $1.148$ & $1.186$ & $3.21\;10^{-4}$ \\
      $1.180$ & $0.515$ & $2.12\;10^{-4}$  \\
      \hline
    \end{tabular}
  \end{minipage}
  \hspace{1.0cm}
  \\
  \\
  \begin{minipage}[b]{1.\linewidth}
    \centering
    \begin{tabular}{|c|c|c|}
      \hline
      $R_{\alpha}$ & $\Phi^{(0)}$ (lbu) & $\sigma_{\mbox{\tiny w}}^{(0)}$ (lbu) \\
      \hline
      $1.000$ & $0.434$  & $1.01\;10^{-4}$   \\
      $1.023$ & $1.297$ & $3.49\;10^{-4}$   \\
      $1.066$ & $1.311$ & $2.86\;10^{-4}$   \\
      $1.107$ & $0.476$  & $1.98\;10^{-4}$   \\
      $1.122$ & $0.503$ & $1.95\;10^{-4}$ \\
      $1.162$ & $0.479$ & $2.04\;10^{-4}$  \\
      \hline
    \end{tabular}
  \end{minipage}
  \caption{Values of $\Phi^{(0)}$ and $\sigma_{\mbox{\tiny w}}^{(0)}$ used to normalize the flow rate $\Phi$ and the wall stress $\sigma_{\mbox{\tiny w}}$ in Fig.~\ref{fig:scalingZoom}.}\label{table}
\end{table}

\begin{figure}[]
\onefigure[scale=0.15]{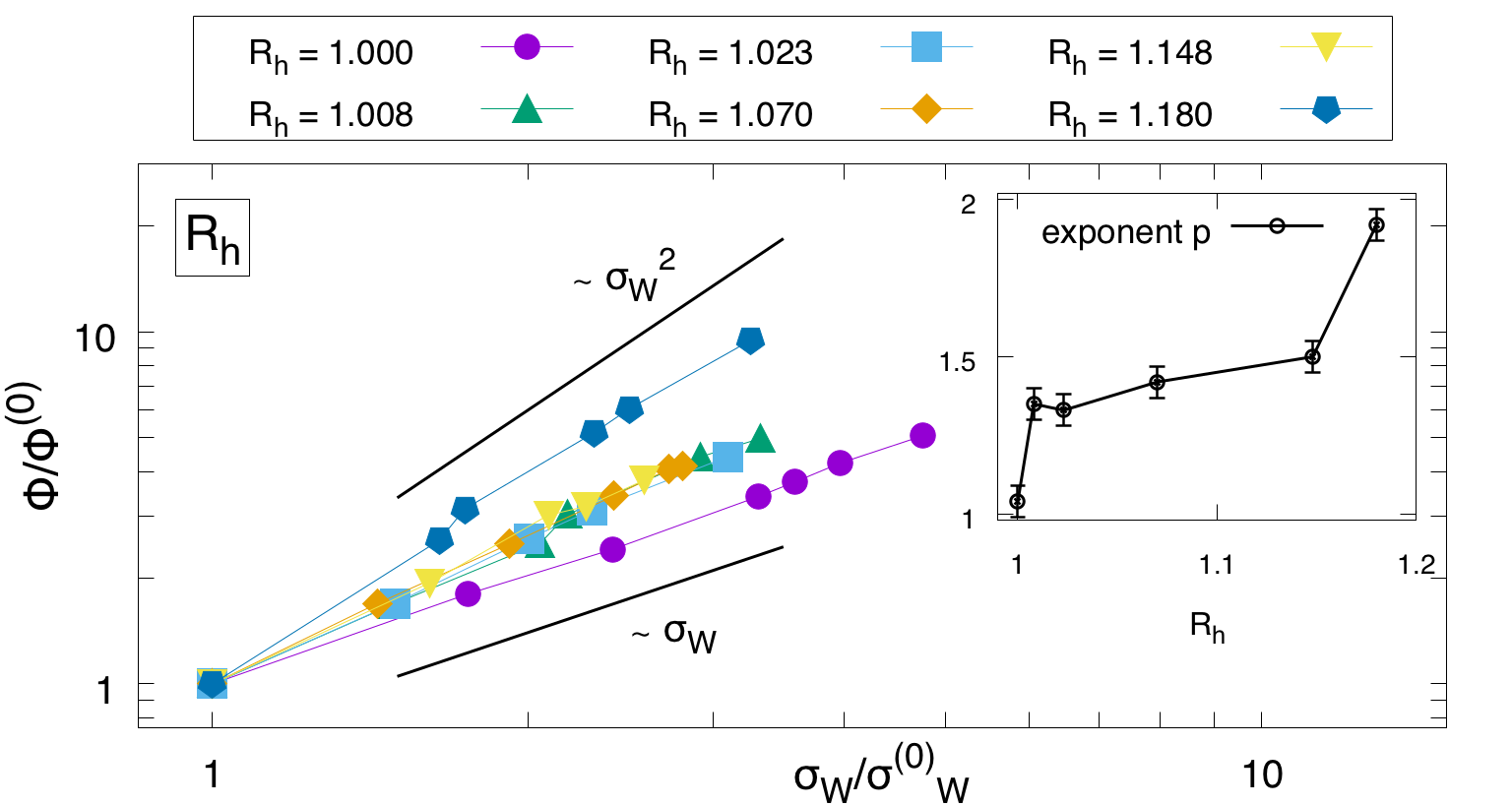}
\onefigure[scale=0.15]{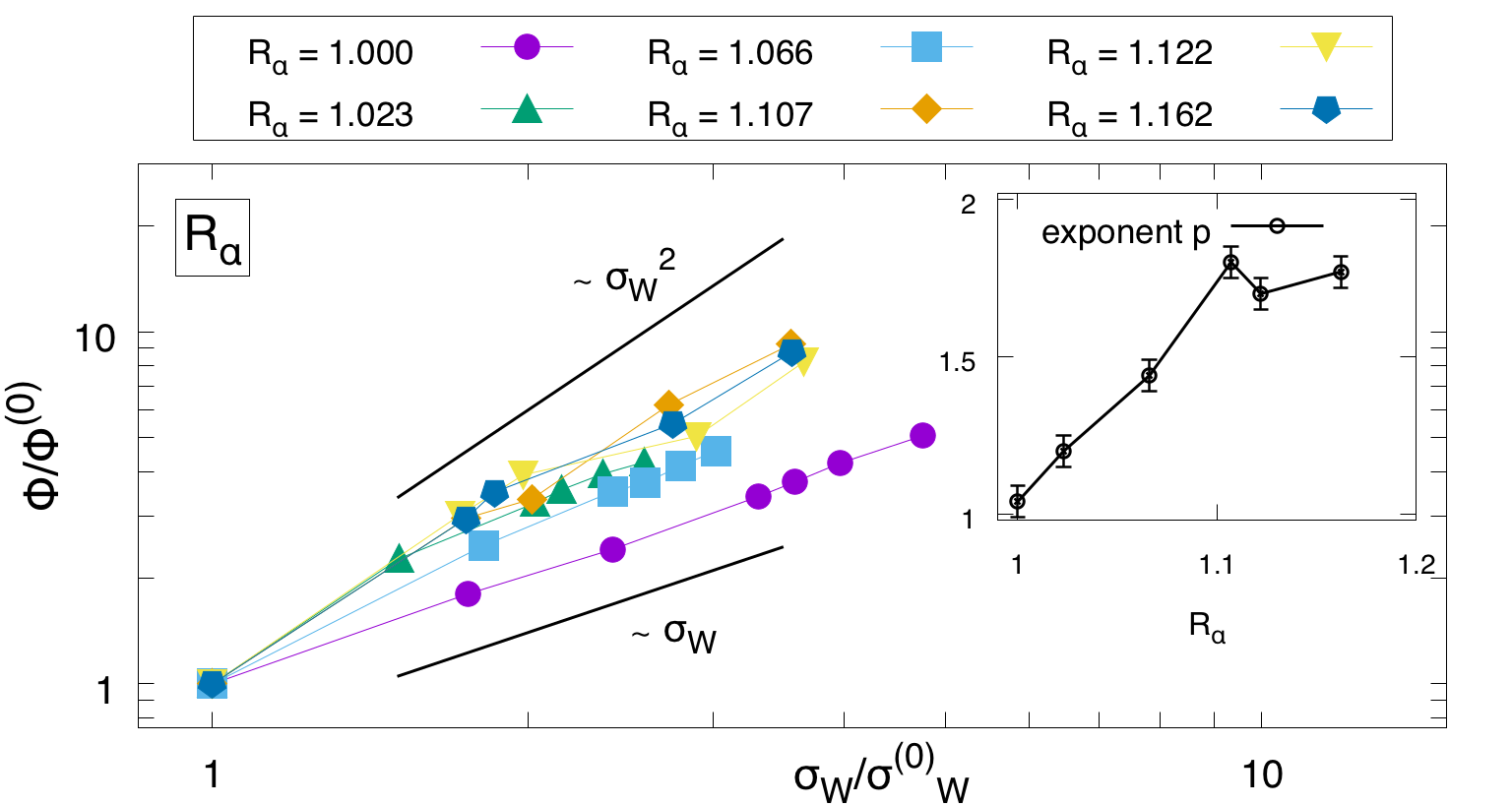}
\caption{Log-log plot of the normalized flow rate $\Phi/\Phi^{(0)}$ as a function of the normalized wall stress 
$\sigma_{\mbox{\tiny w}}/\sigma_{\mbox{\tiny w}}^{(0)}$ for data of Fig.~\ref{fig:scaling}. 
The reference wall stress $\sigma_{\mbox{\tiny w}}^{(0)}$ has been chosen at the beginning of the fluidized region observed 
in Fig.~\ref{fig:scaling} (see also text for details). 
Quantitative values for $\Phi^{(0)}$ and $\sigma_{\mbox{\tiny w}}^{(0)}$ are given in Table \ref{table}. 
Two relevant relations ($\sim \sigma_{\mbox{\tiny w}}$ and $\sim \sigma_{\mbox{\tiny w}}^2$) that are found in the literature 
of soft-materials are drawn for comparison~\cite{Meekeretal04a,Meekeretal04b,Sethetal08,Sethetal12} (see text for details). 
In the insets we report the values of the exponents $p$ relating $\Phi/\Phi^{(0)}$ and 
$\sigma_{\mbox{\tiny w}}/\sigma_{\mbox{\tiny w}}^{(0)}$ via the relation 
$\Phi/\Phi^{(0)}=(\sigma_{\mbox{\tiny w}}/\sigma_{\mbox{\tiny w}}^{(0)})^p$.}
\label{fig:scalingZoom}
\end{figure}
\begin{figure}[t!]
\onefigure[scale=0.15]{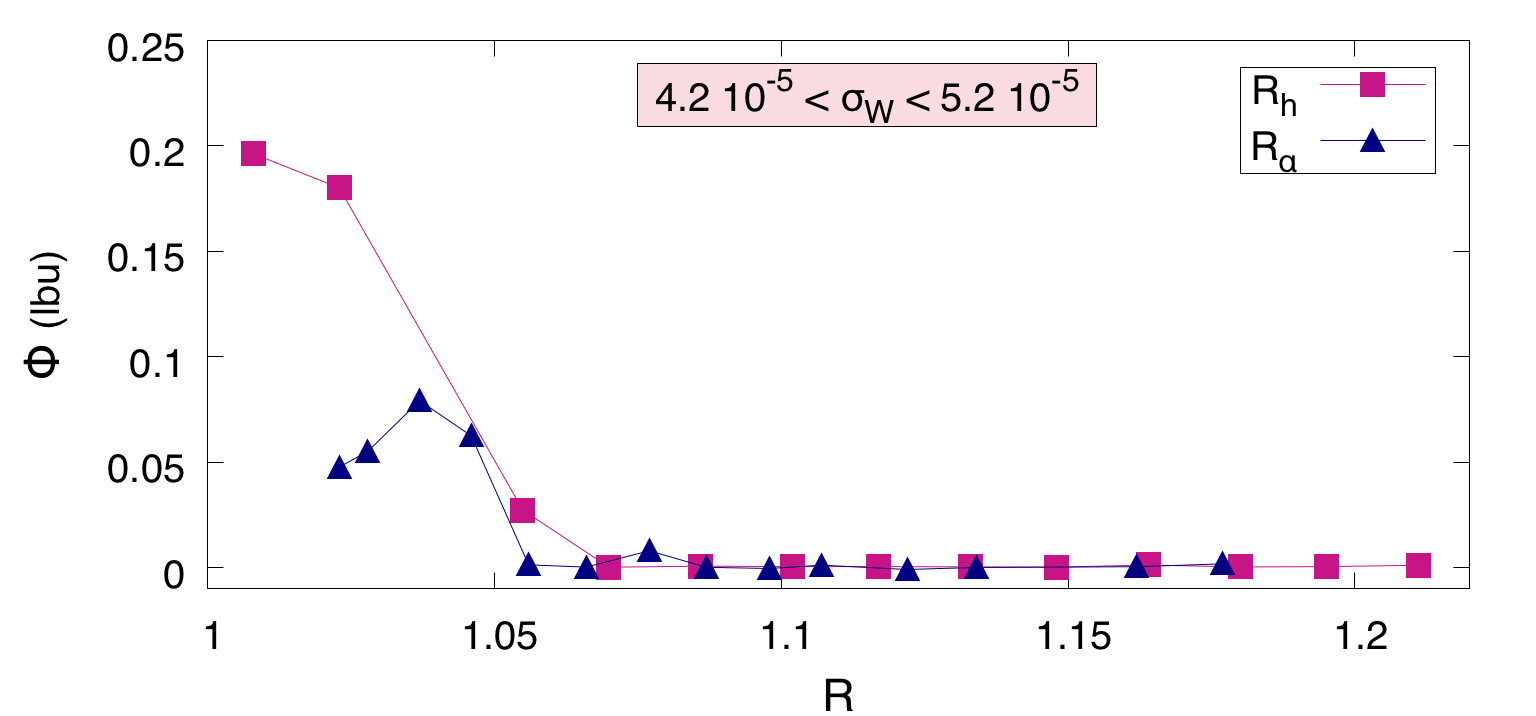}
\onefigure[scale=0.15]{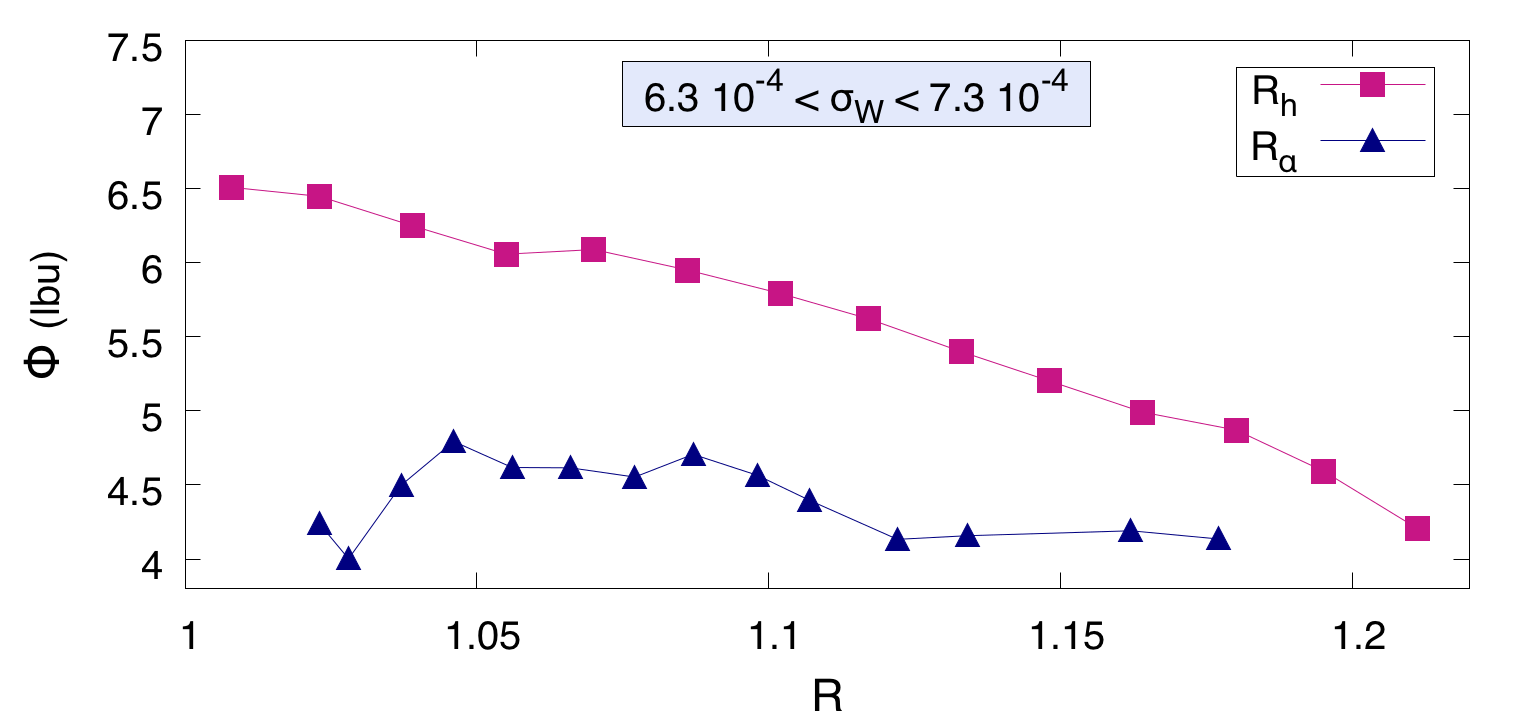}
\caption{Flow rate $\Phi$ as a function of the roughness for the two different roughness shape realizations $R_{\mbox{\tiny h}}$ and $R_{\alpha}$ (see Fig.~\ref{fig:setup}). The value of the wall stress $\sigma_{\mbox{\tiny W}}$ is selected in two intervals (see also shaded areas in Fig.~\ref{fig:scaling}): $\sigma_{\mbox{\tiny W}} \in [4.2\, 10^{-5}: 5.2\, 10^{-5}]$ (``small wall stress'');$\sigma_{\mbox{\tiny W}} \in [6.3\, 10^{-4}: 7.3\, 10^{-4}]$ (``large wall stress'').}
\label{fig:flowRate-R}
\end{figure}
%
\section{Conclusions}
Summarizing, we have systematically analyzed the flow rate $\Phi$ of a soft-material driven by a pressure gradient in a confined 
channel with variable roughness on one wall. To emphasize wall effects, we focused our attention only on values of the rough 
wall stress $\sigma_{\mbox{\tiny W}}$ below -- or of the order of -- the material yield stress, resulting in a flow rate 
dominated by near-wall properties. We have then considered the impact of roughness shape. To that purpose, we have designed two complementary roughness realizations, so as to highlight both the importance of height variations ($R_{\mbox{\tiny h}}$) as well as the roughness localization in space ($R_{\alpha}$). The effect of the roughness results in a suppression of the flow (no-flowing regime) 
for small $\sigma_{\mbox{\tiny W}}$, while a flowing regime sets in at larger $\sigma_{\mbox{\tiny W}}$. 
For values of wall stress just above the inception of the flowing regime, we quantitatively analyzed the relation between 
$\Phi$ and $\sigma_{\mbox{\tiny W}}$, by changing systematically the values and shape of the roughness. 
For smooth walls, it is observed that a linear relation holds, in agreement with previous 
observations~\cite{Meekeretal04a,Meekeretal04b,Sethetal08,Sethetal12,Goyonetal10,Geraudetal13}. 
By increasing the values of the roughness, steeper curves are observed, which have been quantified via the scaling scenario, 
$\Phi \sim \sigma_{\mbox{\tiny W}}^p$, with exponents $p$ larger than 
one~\cite{Meekeretal04a,Meekeretal04b,Sethetal08,Sethetal12,Goyonetal10,Geraudetal13,Mansardetal14,Cloitreetal17}. 
While the existence of such steeper curves generically holds at changing the roughness realizations, 
the steepness has been found to change more gently in the realization $R_{\alpha}$ than in the realization $R_{\mbox{\tiny h}}$. 
Finally, at fixed $\sigma_{\mbox{\tiny W}}$ in the flowing regime, we have also analyzed the dependence of $\Phi$ on the 
roughness value: while for the realization $R_{\alpha}$, $\Phi$ stays practically constant, 
it becomes a decreasing function of the roughness value in the realization $R_{\mbox{\tiny h}}$, 
indicating that the variation in height is mainly responsible for the drop in the flow rate.\\
Taken all together, we argue that these results may be useful for designing microfluidic channels to control and passively drive the motion of a soft-material. For future investigations, multiple pathways are worth being pursued. 
Beyond the precise characterization of the scaling scenario already discussed in the text, 
another important characterization would be that of plasticity. Indeed, roughness inevitably triggers 
rearrangements~\cite{Goyonetal08,Goyonetal10,Mansardetal14,Paredes15,Scagliarinietal16,Derzsietal17,Derzsietal18} 
and could also be interesting to analyze the impact of the roughness shape on such rearrangements. 
Preliminary analysis shows that the flow rate grows with the number of rearrangements~\cite{Derzsietal17,Derzsietal18}, as expected, but the specific details of such dependency may hide a non-trivial dependence on the roughness shape. We also remark that in this work we focused on the time-averaged flow rate, which oscillates weakly in the ``solid'' regime. The flow rate fluctuations tend to grow 
when the transition to fluidization is approached. It is then logical to address how the amplitude and frequencies of these oscillations depend on the wall roughness geometry and how they correlate to the topological characteristics of the droplet assembly constituting the soft-material.

\acknowledgments
FP and MS acknowledge financial support from the project Hydrodynamics of Soft-Glassy materials through microdevices (HYDROSOFT) 
financed by the
University of Rome ``Tor Vergata'' (Mission Sustainability). The research leading to these results has received funding from the European Research Council
under the European Unions Horizon 2020 Framework Programme (No. FP/2014-2020)/ERC Grant Agreement No. 739964 (COPMAT). ML is grateful for the support of Hong Kong GRF (Grant 15330516) and Hong Kong PolyU (Grants 1-ZVGH and G-UAF7).

\end{document}